\documentclass[twocolumn,showpacs,preprintnumbers,amsmath,amssymb]{revtex4}
\usepackage{dcolumn}
\usepackage{bm}
\usepackage{graphicx}%
\usepackage{epsfig}

\begin{document}
\title{Designing reservoirs for $1/t$ decoherence of a qubit}
\author{Filippo Giraldi }
\email{giraldi@ukzn.ac.za,filgi@libero.it}
\affiliation{Quantum Research Group, School of Physics and National Institute for Theoretical Physics, University of KwaZulu-Natal, Durban 4001, South Africa}

\author{ Francesco Petruccione}
\email{petruccione@ukzn.ac.za }
\affiliation{Quantum Research Group, School of Physics and National Institute for Theoretical Physics,
University of KwaZulu-Natal, Durban 4001, South Africa}


\begin{abstract}
Anomalous decoherence in the Jaynes-Cummings model emerges for a certain class of bosonic reservoirs, described by spectral densities with a band edge frequency coinciding with the qubit transition frequency. The special reservoirs are piecewise similar to those usually adopted in Quantum Optics, i.e., sub-ohmic at low frequencies and inverse power laws at high frequencies. The exact dynamics of the qubit is described analytically through Fox $H$-functions. Over estimated long time scales, decoherence results in inverse power laws with powers decreasing continuously to unity, according to the particular choice of the special reservoir. The engineering reservoir approach is a new way of strongly delaying the decoherence process with possible applications to Quantum Technologies, due to the simple form of the designed reservoirs.
\end{abstract}

\maketitle

\section{Introduction}
Decoherence indicates the process that a quantum system of interest undergoes through the interaction with its external environment.
The corresponding time evolution and the destructive effects on quantum coherence are treated in the Theory of Open Quantum Systems \cite{weiss, BP}.

Great attention has been devoted to the dissipative effects of a two-level system (TLS), a qubit in Quantum Information Theory, interacting with an external environment modeled by a reservoir of bosons \cite{WW,legett,JC}. The applications of this simple model are most various: from Nanotechnology to Quantum Information and Quantum Computing, from Quantum Optics to circuit QED, to name a few. Still, the central issue and one of the greatest challenges remains the way to control or delay the destructive effect of the external environment on coherence. For example, the decoherence time \cite{DiVincenzo} in Magnetic Resonance has orders of magnitude ranging between nanoseconds and seconds.

 Various techniques are adopted in order to give an analytical description of the exact dynamics of the qubit. Interesting results emerge from the adoption of the resolvent operator \cite{ctbook} in rotating wave approximation \cite{JC}, with a Lorentzian distribution of field modes \cite{szbook}. The assumption that the coupling constants vary slowly with frequency, allows a complete analytical treatment and the exact dynamics results in oscillating behaviors enveloped in exponential decays. For a detailed report we also refer to \cite{lambropoulos}.

 An interesting model for the spontaneous decay of a TLS in a structured reservoir, has been introduced by Garraway \cite{garraway,garraway2} and solved exactly for a generic discrete or a continuous distribution of field modes, described by Lorentzian type spectral densities and a special non-Lorentzian one with two poles in the lower half plane. The crucial conjecture that the frequency range is $\left(-\infty,+\infty\right)$, allows the analytical evaluation of the exact dynamics in terms of the pseudomodes \cite{garrawayknight} and the poles of the spectral density in the lower half plane. The time evolution results in oscillations enveloped in exponential relaxations. Recently \cite{VB}, the model adopted by Garraway has been chosen to compare the qubit dynamics of an exact master equation in time convolution less form with the Nakajima-Zwanzig master equation, through a perturbation expansion of the memory kernel.

The realization of structured environments providing a discontinuity in the distribution of the frequency modes, also named photonic band gap (PBG) \cite{y87,j87,jpc,jvf}, introduces new phenomena in the atom-cavity interactions.
For example, the spontaneous emission of a two-level atom near the edge of a PBG exhibits oscillatory relaxations \cite{JQ1994} instead of a purely exponential decay.
A theoretical model providing a PBG structure in a $N$-period one dimensional lattice has been proposed in Ref. \cite{bendickson} by arranging an appropriate sequence of the unit lattice cells. The density of the frequency modes is analytically evaluated as a function of the transmission coefficient of each unit cell.

In line with the attempt to contain the destructive effect of the external environment on a qubit, decoherence, a special reservoir of bosons is designed in Ref. \cite{GP} with a PBG structure, excluding modes with frequencies lower than the transition frequency of the qubit. The exact dynamics is described analytically by a linear combination of incomplete Gamma functions and decoherence results in a $3/2$ inverse power law relaxation over an evaluated long time scale. If compared to the exponential like relaxations, decoherence is strongly delayed. In this scenario, we adopt the Jaynes-Cummings model, starting from the initially unentangled states of the system qubit plus reservoir, adopted in Ref. \cite{garraway}, and design a wider class of reservoirs inducing slower forms of relaxation for a further delay the decoherence process. We notice in advance that the analytical calculations leading to the exact dynamics, rest on a continuous distribution of positive mode frequencies.

\section{The model}

The interaction between the qubit and the dissipative environment is described through the Jaynes-Cumming model with a continuous distributions of field modes \cite{JC, BP,weiss, garraway,VB}. By choosing $\hbar=1$, the
Hamiltonian of the whole system is $H_S+H_E+H_I$, where
\begin{eqnarray}
H_S=\omega_0 \,\sigma_+\sigma_-,
\hspace{1em}H_E=\sum_{k=1}^{\infty} \omega_k \, a^{\dagger}_k a_k, \nonumber \\
H_I=
\sum_{k=1}^{\infty} \left(g_k \, \sigma_+\otimes a_k+g_k^{\ast}\,
\sigma_-\otimes a^{\dagger}_k\right).\nonumber
\end{eqnarray}
 The rising and lowering operators, $\sigma_+$ and $\sigma_-$, respectively, act on the Hilbert space of the qubit, defined through the equalities
  $\sigma_+=\sigma_-^{\dagger}=|1\rangle\langle0|$,
   while $a_k^{\dagger}$ and $a_k$ are the creation and annihilation operators,
   respectively, acting on the Hilbert space of the $k$-th boson, fulfilling the commutation rule
   $\left[a_k,a_k^{\prime^{\dagger}}\right]=\delta_{k,k^{\prime}}$ for every
   $k,k^{\prime}=1,2,3,\ldots$.
   The constants $g_k$ represent the coupling between the transition $|0\rangle \leftrightarrow |1\rangle$ and the $k$-th mode of the radiation field, while $\omega_0$ is the qubit transition frequency.
   In the following we refer to the system of a TLS interacting with a cavity supplying a reservoir of field modes, as studied by Garraway \cite{garraway} and adopted in Ref. \cite{VB}. Starting from the initial state of the total system
   \begin{equation}
|\Psi(0)\rangle=\left(c_0 |0\rangle+ c_1(0)|1\rangle\right) \otimes |0\rangle_E, \label{Psi0}
\end{equation}
   where $|0\rangle_E$ is the vacuum state of the environment, the exact time evolution is
   described by the form
   \begin{eqnarray}
   &&|\Psi(t)\rangle=c_0 |0\rangle \otimes |0\rangle_E+c_1(t)|1\rangle \otimes |0\rangle_E \nonumber\\
   &&+ \sum_{k=1}^{\infty}b_k(t)|0\rangle \otimes |k\rangle_E,\hspace{0.5em}|k\rangle_E=a^{\dagger}_k |0\rangle_E, \hspace{0.5em}k=0,1,2,\ldots. \nonumber
\end{eqnarray}
  The dynamics is easily studied in the interaction picture,
   \begin{eqnarray}
   &&|\Psi(t)\rangle_I=e^{\imath \left(H_S+H_E\right)t}|\Psi(t)\rangle =c_0 |0\rangle \otimes |0\rangle_E \nonumber \\
   &&+\,C_1(t)|1\rangle\otimes|0\rangle_E
+\sum_{k=1}^{\infty}B_k(t)|0\rangle \otimes |k\rangle_E, \nonumber
\end{eqnarray}
where $\imath$ is the imaginary unity, $C_1(t)=e^{\imath \omega_0 t}\,c_1(t)$ and $B_k(t)=e^{\imath \omega_k t}\,b_k(t)$ for every $k=1,2,\ldots$.
The Schr\"odinger equation gives the forms:
\begin{eqnarray}
&&\dot{C}_1(t)=-\imath \sum_{k=1}^{\infty} g_k \, B_k(t)\,e^{-\imath\left(\omega_k-\omega_0\right)t}, \nonumber \\
&& \dot{B}_k(t)=-\imath \,g^{\ast}_k \, C_1(t)\,e^{\imath\left(\omega_k-\omega_0\right)t}, \nonumber
\end{eqnarray}
leading to the following convoluted structure equation for the amplitude $\langle 1| \otimes\, _E\langle 0|| \Psi(t) \rangle_I$, labeled as $C_1(t)$,
\begin{equation}
\dot{C}_1(t)=-\left(f\ast C_1\right)(t),
\label{cMEQ&CorrReservoir}
\end{equation}
 where  $f$ is the two-point correlation function of the reservoir of field modes, \begin{eqnarray}
f\left(t-t^{\prime}\right)=\sum_{k=1}^{\infty}
\left|g_k\right|^2 e^{-\imath \left(\omega_k-\omega_0\right)
\left(t-t^{\prime}\right)}. \nonumber
\end{eqnarray}
  For a continuous distribution of modes described by  $\eta\left(\omega\right)$, the correlation function is expressed through the spectral density function $J\left(\omega\right)$,
 \begin{eqnarray}
f\left(\tau\right)=
\int_0^{\infty}J\left(\omega\right) e^{-\imath\left(\omega-\omega_0\right)\tau }
d \omega, \nonumber
\end{eqnarray}
where $J\left(\omega\right)=\eta\left(\omega\right) \left|g\left(\omega\right)\right|^2$ and $g\left(\omega\right)$ is the frequency dependent coupling constant.

The exact dynamics of the qubit is described by the time evolution of the reduced density matrix obtained by tracing over the Hilbert space of the bosons,
\begin{eqnarray}
&&\rho_{1,1}(t)=1-\rho_{0,0}(t)=\rho_{1,1}(0)\,\left|G(t)\right|^2,
\label{rhot11}\\
&&\rho_{1,0}(t)=\rho_{0,1}^{\ast}(t)=\rho_{1,0}(0)\,e^{-\imath \omega_0 t  }G(t).
\label{rhot10}
\end{eqnarray}
The function $G(t)$, fulfilling the convolution equation
\begin{equation}
\dot{G}(t)=-\left(f\ast G\right)(t),\hspace{1em}G(0)=1.
\label{G}
\end{equation}
The function drives both the dynamics of the levels populations and the decoherence term.

\section{The exact dynamics}

We study the exact dynamics of the reduced density matrix of the qubit, interacting in rotating wave approximation with a reservoir of bosons described by the continuous spectral density
\begin{eqnarray}
&&J_{\alpha}\left(\omega\right)=
 \frac{2 A \left(\omega-\omega_0\right)^{\alpha}
 \Theta\left(\omega-\omega_0\right)}
 {a^2+\left(\omega-\omega_0\right)^2 },\label{Ja} \\
&& A>0,\hspace{1em}a>0, \hspace{1em}1>\alpha>0.\nonumber
\end{eqnarray}
This simple form exhibits a PBG edge in the qubit transition frequency, has an absolute maximum $M_{\alpha}$ at the frequency $\Omega_{\alpha}$,
 \begin{eqnarray}
 &&M_{\alpha}=J_{\alpha}\left(\Omega_{\alpha}\right)=A\, \alpha^{\alpha/2} a^{\alpha-2}\left(2-\alpha\right)^{1-\alpha/2}, \nonumber \\&&\Omega_{\alpha}=\omega_0+a \,\alpha^{1/2}\left(2-\alpha\right)^{1/2}.\nonumber
 \end{eqnarray}
The above spectral densities are piece-wise similar to those usually adopted, i.e. sub-ohmic at low frequencies, $\omega \simeq \omega_0$, and inverse power laws at high frequencies, $\omega \gg \omega_0$, similar to the Lorentzian one, though with different power,
\begin{eqnarray}
&&J_{\alpha}\left(\omega\right)\sim
 2 A/a^2 \left(\omega-\omega_0\right)^{\alpha}, \hspace{2em}
 \omega \to \omega_0^+,
 \nonumber \\
 && J_{\alpha}\left(\omega\right)\sim
 2 A \,\omega^{\alpha-2}, \hspace{2em} \omega \to +\infty. \nonumber
 \end{eqnarray}

 The \emph{exact} dynamics of a qubit interacting with a reservoir of bosons described by the spectral density $J_{\alpha}\left(\omega\right)$, is driven by the function $G_{\alpha}(t)$, solution of Eq. (\ref{G}),
 \begin{eqnarray}
     G_{\alpha}(t)=\sum_{n=0}^{\infty}\sum_{k=0}^n\frac{(-1)^n\, n!\,z_{\alpha}^k \,z_0^{n-k}\,t^{3n-\alpha k}}{k!(n-k)!} \nonumber \\
     \times\Big(E^{n+1}_{2,3n-\alpha k+1}\left(-z_1 t^2\right)
     -a^2 t^2  E^{n+1}_{2,3n-\alpha k+3}\left(-z_1 t^2\right)\Big), \label{GaMittag}
    \end{eqnarray}
   expressed as a series of Generalized Mittag-Leffler functions \cite{Prabhakar,MainardiBook},
  \begin{eqnarray}
    &&E_{\alpha,\beta}^{\gamma}\left(z\right)=\sum_{n=0}^{\infty}
  \frac{\left(\gamma\right)_n z^n}{n!\, \Gamma\left(\alpha n+\beta\right)}, \nonumber\\
  &&\alpha, \beta, \gamma\in C,\hspace{2em} \Re\left\{\alpha\right\}>0,\hspace{1em}\Re\left\{\beta\right\}>0,\nonumber
  \end{eqnarray}
  where $\left(\gamma\right)_0=1$ and $\left(\gamma\right)_n=\Gamma\left(\gamma+n\right)/\Gamma\left(\gamma\right)$.
  The parameters involved read
    \begin{eqnarray}
    &&z_1=\pi A a^{\alpha-1}\sec\left(\pi \alpha /2\right)-a^2, \, z_0=\imath \pi A a^{\alpha} \cos\left(\pi \alpha/2\right),\nonumber
\\&&z_{\alpha}=-2\imath \pi A e^{-\imath \pi \alpha/2} \csc\left(\pi \alpha\right).\label{parameters}
\end{eqnarray}
The proof is performed below.

  The Generalized Mittag-Leffler function, fundamental in Fractional Calculus \cite{MainardiBook}, is a particular case of the Fox $H$-function, defined through a Mellin-Barnes type integral in the complex domain,
  \begin{eqnarray}
H_{p,q}^{m,n}\left[z\Bigg|
\begin{array}{rr}
\left(a_1,\alpha_1\right), \ldots,\left(a_p,\alpha_p\right)\\
\left(b_1,\beta_1\right), \ldots,\left(b_q,\beta_q\right)\,\,
\end{array}
\right]
=\frac{1}{2 \pi \imath} \nonumber \\
\times\int_{\mathcal{C}}
\frac{\Pi_{j=1}^m \Gamma\left(b_j+\beta_j s\right)
\Pi_{m=1}^n \Gamma\left(1-a_l-\alpha_l s\right)
z^{-s}}{\Pi_{l=n+1}^p \Gamma\left(a_l+\alpha_l s\right)
\Pi_{j=m+1}^q \Gamma\left(1-b_j-\beta_j s\right)
} \,  ds.\nonumber
\end{eqnarray}
Under the conditions that the poles of the Gamma functions in the dominator, do not coincide, also the empty products are interpreted as unity. The natural numbers $m,n,p,q$ fulfill the constraints: $0\leq m\leq q$, $0\leq n\leq p$, and $\alpha_i,\beta_j\in \left(0,+\infty\right)$ for every $i=1,\cdots,p$ and $j=1,\cdots,q$. For the sake of shortness, we refer to \cite{HbookMSH} for details on the contour path $\mathcal{C}$, the existence and the properties of the Fox $H$-functions.
The relation
\begin{eqnarray}
E_{\alpha,\beta}^{\gamma}(-z)=\frac{1}{\Gamma\left(\gamma\right)}
\,H_{1,2}^{1,1}\left[z\Bigg| \begin{array}{rr}
\left(1-\gamma,1\right) \hspace{3em}
\\
\left(0,1\right), \left(1-\beta,\alpha\right)
\end{array}
\right], \nonumber
\end{eqnarray}
leads to a series solution of Fox $H$ functions,
\begin{eqnarray}
    G_{\alpha}(t)=\sum_{n=0}^{\infty}\sum_{k=0}^n\frac{(-1)^n\, z_{\alpha}^k \,z_0^{n-k}\,t^{3n-\alpha k}}{k!(n-k)!} \nonumber \\
      \times\Bigg(H_{1,2}^{1,1}\left[z_1 t^2\Bigg|
      \begin{array}{rr}
\left(-n,1\right) \hspace{5em}
\\
\left(0,1\right), \left(\alpha k-3 n,2\right)
\end{array}
\right] \nonumber \\
     -\,a^2 t^2 H_{1,2}^{1,1}\left[z_1 t^2\Bigg| \begin{array}{rr}
\left(-n,1\right)\hspace{6.7em}
\\
\left(0,1\right), \left(\alpha k-3 n-2,2\right)
\end{array}
\right]\Bigg)
     . \label{GaH}
    \end{eqnarray}
The Generalized Hypergeometric, the Wright \cite{WrightKST} and the Meijer $G$-functions \cite{Erdely1} are particular cases of the Fox $H$-function, thus, $G_{\alpha}(t)$ can be expressed as a series of each of these Special functions, as well.

Particular cases give simplified solutions. For example, the condition $A=A^{\left(\star\right)}$,
\begin{equation}
A^{\left(\star\right)}= \frac{a^{3-\alpha}}{\pi}\, \cos\left(\pi \alpha/2\right),
\end{equation}
 corresponding to $z_1=0$, gives a power series solution,
 \begin{eqnarray}
          G_{\alpha}^{\left(\star\right)}(t)
 = \sum_{n=0}^{\infty}\sum_{k=0}^n\frac{(-1)^n\, n!\, z_{\alpha}^k \,z_0^{n-k}\,t^{3n-\alpha k}}{k!\,(n-k)!\,\Gamma\left(3n-\alpha k+1\right)}
 \nonumber \\ \times
  \Bigg\{ 1-a^2\,\frac{ \,\Gamma\left(3n-\alpha k+1\right)\, }{\Gamma\left(3n-\alpha k+3\right)}\,t^2 \Bigg\}, \hspace{1em} 1>\alpha>0. \label{Ghyperz10}
        \end{eqnarray}

If the parameter $\alpha$ takes rational values, $p/q$, where $p$ and $q$ are distinct prime numbers such that $0<p<q$, the solution of Eq. (\ref{G}) can be expressed as a modulation of exponential relaxations \cite{NM},
\begin{equation}
 G_{p/q}(t)=\int_0^{\infty}d\eta \,\int_0^{\infty}d\xi\, \Phi_{p/q}\left(\eta, \xi\right) e^{-\xi t},
 \label{GpqtI}
 \end{equation}
 where
 \begin{eqnarray}
 &&\Phi_{p/q}\left(\eta, \xi\right)=\sum_{l=1}^n\sum_{k=1}^{m_l} \frac{b_{l,k}\left(\zeta_l\right)}{\pi}\,
 \, \eta^{m_l-k} \nonumber \\
 &&\times\sin\left(\eta\, \xi^{1/q}\sin \left(\pi/q\right)\right)e^{\eta\left(\zeta_l- \cos\left(\pi/q\right)\xi^{1/q}\right)}. \nonumber
  \end{eqnarray}
   The rational functions $b_{l,k}\left(z\right)$ read
  \begin{eqnarray}
  b_{l,k}\left(z\right)=\frac{d^{k-1}/dz^{k-1}\,\left[\left(z^q-a\right)
  \left(z^q+a\right)\left(z-\zeta_l\right)^{m_l}/ Q\left(z\right)\right]}
  { \left(m_l-k\right)!\left(k-1\right)!}, \nonumber
   \end{eqnarray}
   for every $l=1,\cdots,n$, and $k=1,\ldots ,m_l$. The complex numbers $\zeta_1, \cdots, \zeta_n$ are the roots of the polynomial
 \begin{equation}
 Q(z)=z^{3q}+z_1\, z^q+z_{\alpha}\,z^{p}+z_0 \label{Q}
 \end{equation}
  and $m_l$ is the multiplicity of $\zeta_l$, for every $l=1,\ldots,n$, which means $Q(z)=\Pi_{l=1}^n \left(z-\zeta_l\right)^{m_l}$ and $\sum_{l=1}^{n}m_l=3 q$.

  The case $\alpha=1/2$ exhibits a simplified exact dynamics described by a finite sum of Eulerian functions \cite{GP}. For $\alpha=3/4$ and $A=a^{9/4}\cos\left(3 \pi/8\right)/ \pi$, the parameter $z_1$ vanishes and the roots $\zeta_1,\ldots,\zeta_l$ can be evaluated analytically from the solutions of a quartic equation. We do not report the expressions for the sake of shortness. In the remaining cases of rational values of $\alpha$, the roots of $Q(z)$ must be evaluated numerically, once the numerical values of both $A$ and $a$ are fixed.
  These details complete the necessary analysis of the function $G_{\alpha}(t)$.

 Finally, the exact time evolution of the qubit is obtained from Eqs. (\ref{rhot11}) and (\ref{rhot10}), by replacing the function $G(t)$ with $G_{\alpha}(t)$, analyzed in the present Section.

\section{Inverse power laws}
The theoretical analysis of the \emph{exact} dynamics, performed above, leads to the following concrete result: a \emph{time scale} $\tau$
  emerges such that, for $t\gg \tau$, the function $G_{\alpha}(t)$ exhibits inverse power law behavior described by the asymptotic form
\begin{equation}
G_{\alpha}(t)\sim
   -\mathcal{D}_{\alpha} \, t^{-1-\alpha},\hspace{1em} t\to +\infty,
   \hspace{1em} 1>\alpha>0,\label{Gasympt}
   \end{equation}
   where
   \begin{equation}
   \mathcal{D}_{\alpha}= \frac{2 \,\imath\,\alpha\, a^{2\left(1-\alpha\right)}e^{-\imath \pi \alpha /2} \csc\left(\pi \alpha\right)\sec^2\left(\pi \alpha/2\right)}{\pi A\, \Gamma\left(1-\alpha\right)}. \nonumber
      \end{equation}
   A simple choice is
    \begin{equation}
\tau=\max\left\{1,\left|\frac{3}{z_0}\right|^{1/3},
\left|\,3\,\frac{z_{\alpha}}{z_0}\right|^{1/\alpha},
3\left|\frac{z_1}{z_0}\right| \right\},
 \label{tau}
\end{equation}
   the proof is performed below.

  Thus, over long timescales, $t\gg \tau$, the qubit exact dynamics is described by inverse power law relaxations:
   \begin{eqnarray}
 &&\rho_{1,1}(t)=1-\rho_{0,0}(t)\sim\rho_{1,1}(0)\,
\left|\mathcal{D}_{\alpha}\right|^2 t^{-2-2\alpha},\label{rhot11asympt}
\\
&&\rho_{1,0}(t)=\rho_{0,1}^{\ast}(t)=\rho_{1,0}(0)\,\mathcal{D}_{\alpha}\,e^{-\imath \omega_0 t  } \, t^{-1-\alpha}, \label{rhot10asympt}
\end{eqnarray}
for every $\alpha\in\left(0,1\right)$.

 In Ref. \cite{GP} the decoherence processes corresponding to a Lorentzian and $J_{1/2}\left(\omega\right)$ spectral densities are compared. The exponential type relaxations emerging in  the Lorentzian case, vanish faster than the inverse power laws related to the spectral densities  $J_{\alpha}\left(\omega\right)$, for every $\alpha \in (0,1)$.

 The results obtained are summarized as follows. Starting from the initial condition (\ref{Psi0}) where the reservoir, in the vacuum states, and the qubit are unentangled, we evaluate analytically the exact dynamics of the qubit interacting, in a rotating wave approximation with a reservoir of bosons described by the spectral density (\ref{Ja}). The time evolution is expressed by the series  (\ref{GaH}) of Fox $H$-functions, and, over long timescale, $t\gg \tau$, decoherence results in an inverse power law relaxation proportional to $t^{-1-\alpha }$ for every $\alpha \in (0,1)$, according to the choice of the special reservoir (\ref{Ja}). The qubit ultimately collapses into the ground state.

\section{Conclusions}

Anomalous forms of qubit decoherence emerge from the
Jaynes-Cummings model for reservoirs of bosons described
by special continuous spectral densities with a PBG edge coinciding with the qubit transition frequency. The designed spectral densities are piece-wise similar to those usually adopted, i.e., sub-ohmic at low frequencies and inverse power laws at high frequencies, similar to the Lorentzian one. Initially, the qubit and the reservoir, versing in the vacuum state, are unentangled. The exact dynamics is described analytically through series of Fox $H$-functions. Over estimated long time scales, qubit decoherence results in inverse power law relaxations with powers decreasing continuously to \emph{unity}, according to the choice of the special reservoir.

An environment supplying the designed reservoir of field modes can in principle be realized with materials providing the PBG structure.
For example, a $N$-period one dimensional lattice can reproduce a band gap by arranging the appropriate sequence of dielectric unit cells. The corresponding density of frequency modes is structured by the transmission coefficients of each unit cell. Also, the advanced technologies concerning diffractive grating and photonic crystals allows the realization of tunable 1D PBG microcavities \cite{PBGMC1,PBGMC2}. The simple form of the designed reservoir may be accessible experimentally. The action of such a structured environment on a qubit could be a way of delaying the decoherence process with fundamental applications to Quantum Information processing Technologies.

\section{Proof}
A detailed demonstration of the solutions driving the exact dynamics follows.
The reservoirs of bosons described by the following class of non negative, non divergent and summable spectral densities are considered:
\begin{equation}
  \int_0^{\infty}J\left(\omega\right)
d \omega<\infty,\hspace{1em} J\left(\omega\right)=\Theta\left(\omega-\omega_0\right)
\Lambda\left(\omega-\omega_0\right),\label{Jconstraint}
\end{equation}
in this way, Eq. (\ref{G}) gives
\begin{eqnarray}
&&\tilde{G}(u)=\left\{u-\imath\,\mathcal{S}\left(\Lambda\right)
\left(-\imath u\right)\right\}^{-1}, \label{Gu}\\
 &&\Re \left\{u\right\}>0, \hspace{1em}
\left|\arg\left\{-\imath u\right\}\right|<\pi,
\nonumber
\end{eqnarray}
where $\mathcal{S}$ is the Stieltjes transform.
The constraints:
$\int_0^{\infty} \Lambda\left(\omega\right)d \omega<\infty$, and
$\Re\left\{u\right\}>0$, guarantee the uniform convergence of integrals involved in the Integral transforms, so that the above equality holds true.
The class of spectral densities (\ref{Ja}) and Eq. (\ref{G}) lead to the following Laplace transform:
\begin{equation}
\tilde{G}_{\alpha}(u)=\frac{u^2-a^2}{u^3+ z_1 u
+z_{\alpha}u^{\alpha}+z_0},
\label{Gu}
\end{equation}
the parameters $z_0$, $z_{\alpha}$ and $z_1$ are defined by Eq. (\ref{parameters}).

The function $G_{\alpha}(t)$ is obtained through the \emph{convergent}
 term by term Laplace inversion of the series expansion of Eq. (\ref{Gu}),
 \begin{eqnarray}
    &&\frac{u^2-a^2}{u^3+z_1 u+z_{\alpha} u^{\alpha}+z_0}=\sum_{n=0}^{\infty}\sum_{k=0}^n\frac{(-1)^n\, n!\, z_{\alpha}^k \,z_0^{n-k}}{k!(n-k)!}\nonumber\\
    &&\times \,\frac{u^{\alpha k-n-1}(u^2-a^2)}{\left(u^2+z_1\right)^{n+1}},\hspace{2em}
   \left| \frac{z_{\alpha}u^{\alpha}+z_0}{u^3+z_1 u}\right|<1.
     \nonumber
    \end{eqnarray}
This way, a series solution of Eq. (\ref{G}) can be built in terms of either Generalized Mittag-Leffler functions, Eq. (\ref{GaMittag}), or Fox $H$ functions, Eq. (\ref{GaH}), for details we refer to \cite{Prabhakar,HbookMSH}.

The inverse power law behavior of $G_{\alpha}(t)$ over long time scales, is found through the series expansion
\begin{eqnarray}
    &&\frac{u^2-a^2}{u^3+z_1 u+z_{\alpha} u^{\alpha}+z_0}=\sum_{n=0}^{\infty}\sum_{k=0}^n \sum_{j=0}^k
    \frac{(-1)^n\, n!}{j!(n-k)!(n-j)!} \nonumber\\
    &&\times\,
    z_0^{-n-1} z_{\alpha}^{n-k} z_1^{k-j}(u^2-a^2)\,u^{\alpha (n-k)+k+2j},
    \label{Gu0}
        \end{eqnarray}
    holding true under the constraint
    \begin{equation}
    \left|\frac{u^3+z_1 u+z_{\alpha} u^{\alpha}}{z_0}\right|<1.
        \label{costraintTau}
\end{equation}
The formal term by term Laplace inversion of Eq. (\ref{Gu0}) gives
\begin{eqnarray}
G_{\alpha}(t)\sim  \sum_{n=1}^{\infty}\sum_{k=0}^n \sum_{j=0}^k
    \frac{(-1)^{n+1}\, n!\,z_0^{-n-1}\,z_{\alpha}^{n-k}\, z_1^{k-j}}{j!\,(n-k)!\,(n-j)!}\nonumber
    \\
    \times    \left(a^2\,\frac{\Gamma\left(\alpha(k-n)-k-2 j-2\right)}
    {\Gamma\left(\alpha(k-n)-k-2 j\right)}-t^{-2}\right) \nonumber
     \\ \times \,\frac{t^{-1-\alpha(n-k)-k-2j}}
    {\Gamma\left(\alpha(k-n)-k-2 j-2\right)},\hspace{1em} t\to +\infty,
  \label{InvPowerSeriesGa}
  \end{eqnarray}
    leading to the asymptotic solution (\ref{Gasympt}).
The time scale for inverse power law behavior descends from the inequality (\ref{costraintTau}) and the constraint $t\gg 1$, requested by Eq. (\ref{InvPowerSeriesGa}). A possible choice is obtained by imposing that the absolute value of each term of the left side of the inequality (\ref{costraintTau}) is less than $1/3$. This way, an upper bound for $\left|u\right|$ is obtained and the corresponding inverse estimates a time scale for inverse power laws, given by Eq. (\ref{tau}).

\begin{acknowledgments}
This work is based upon research supported by the South African Research Chairs Initiative of the Department of Science and Technology and National Research Foundation. F.G. is deeply grateful to Prof. R. Nigmatullin for the useful discussions, to Dr. J. Riccardi for the hints about the numerical check and to Prof. F. Mainardi for the continued suggestions on Integral Transforms and Special Functions over the years.
\end{acknowledgments}


\begin{thebibliography}{0}
 \bibitem{BP} H.P. Breuer and F. Petruccione, \emph{The Theory of Open Quantum Systems},
 Oxford University Press, Oxford (2002).
  \bibitem{weiss} U. Weiss, \emph{Quantum Dissipative systems},
3rd ed. World Scientific, Singapore (2008).
 \bibitem{WW}
  V. Weisskopf and E. Wigner,
  Z. Phys., Vol. { \bf 63} pp. 54-73, (1930).
 \bibitem{legett}A.J. Leggett, S. Chakravarty,
A.T. Dorsey, M.P.A. Fisher, A. Garg and W. Zwerger, Rev. Mod. Phys.,
Vol. {\bf 59}, No. 1 (1987).
\bibitem{JC}  E.T. Jaynes  and F.W. Cummings , Proc. \emph{IEEE} Vol. {\bf51}, 89 (1963).

\bibitem{DiVincenzo}D.P. DiVincenzo, Science Vol.{ \bf 270} n. 5234, pp. 255-261 (1995).


    \bibitem{ctbook}  C. Cohen-Tannoudji, J.  Dupont-Roc  and  G. Grynberg, \emph{Atom-Photon Interactions}, Wiley, New York (2004).


\bibitem{szbook} M.O. Scully  and M.S. Zubairy,  \emph{Quantum Optics}, Cambridge University Press, Cambridge (1997).

 \bibitem{lambropoulos} P. Lambropoulos, G.M. Nikolopoulos, T.R. Nielsen and S. Bay, Rep. Prog. Phys. {\bf 63},  455-503 (2000).



    \bibitem{garraway} B.M. Garraway, Phys. Rev. A {\bf55}, 2290 (1997).
\bibitem{garraway2} B.M. Garraway, Phys. Rev. A {\bf55}, 4636 (1997).

    \bibitem{garrawayknight} B.M. Garraway and P.L. Knight, Phys. Rev. A {\bf54}, 3592 (1996).

       \bibitem{VB} B. Vacchini and H. Breuer, Phys. Rev. A {\bf 81}, 042103 (2010).

     \bibitem{y87}Y. Yablonovitch, Phys. Rev. Lett. {\bf 58}, 2059 (1987).

        \bibitem{j87} S. John, Phys. Rev. Lett. {\bf 58}, 2486 (1987).

   \bibitem{jpc} J.D. Joannopoulos, \emph{Photonic Crystals: Molding the Flow of Light}, Princeton, NJ: Princeton University Press (1994).
        \bibitem{jvf}J.D. Joannopoulos, P.R. Villeneuve  and S. Fan,
       Nature,  {\bf 386},  143-149 (1997).

        \bibitem{JQ1994} S. John and  T. Quang, Phys. Rev. A {\bf 50}, 1764-1769 (1994).

    \bibitem{bendickson} J.M. Bendickson, J.P. Dowling and M.  Scalora, Phys. Rev. E {\bf 53}, 4107 (1996).

     \bibitem{GP}  F. Giraldi and F. Petruccione,  arXiv:1011.0059.
  \bibitem{MainardiBook} F. Mainardi, \emph{Fractional Calcolus and Waves in Linear Viscoelasticity}, Imperial College Press, World Scientific Publishing (2010).




 \bibitem{Prabhakar}T.R.  Prabhakar, Yokohama Mathematical Journal {\ bf 19}, 7  (1971).

 \bibitem{HbookMSH} A.A.  Mathai,  R.K. Saxena and H.J. Haubold \emph{The H-Function, Theory and Applications},  Springer (2009).
 \bibitem{WrightKST}  Kilbas A.K., Saigo M. and  Trujillo J.J., Fract. Calc. Appl. Anal., { \bf 5}: 4, 437-460 (2002).

     \bibitem{em1}A. Erdélyi, W. Magnus, F. Oberhettinger F. and F.G. Tricomi, \emph{Higher Trascendental Functions} Vol. I, McGraw-Hill, New York (1953).

     \bibitem{NM}   R.R. Nigmatullin and F. Mainardi, {\bf IMACS '94},
Proceeding of the 14th IMACS World Congress on Computational and Applied Mathematics, Vol. 1, pp. 370-374 (1994).

\bibitem{PBGMC1} C.W. Wong, X. Yang, P.T. Rackic, S.G. Johnson,M. Qi,Y. Jeon, G. Barbastathis and S. Kim, Appl. Phys. Lett. Vol. {\bf 84}, 1242 (2004).
    
\bibitem{PBGMC2}J.S. Foresi, P.R. Villeneuve, J. Ferrera, E.R. Thoen, G. Steinmeyer, S. Fan, J.D. Joannopoulos, L.C. Kimerling, H.I. Smith and E. P. Ippen, Nature {\bf 390}, 143 (1997).


\end{thebibliography}
\end{document}